\documentclass[preprint,preprintnumbers,amsmath,amssymb]{revtex4}
\usepackage{graphicx}% Include figure files
\usepackage{dcolumn}% Align table columns on decimal point
\usepackage{bm}% bold math
\usepackage{epsfig}

\def\integral{\int_{-\infty}^{+\infty}}
\def\pre{\frac{1}{\sqrt{2\pi\hbar}}}
\begin{document}

\preprint{\bf PSU/TH-254} %\hskip 1cm {\bf May 2004}}

\title{Analytic results for Gaussian wave packets in four model systems:
\vskip 0.1cm
II. Autocorrelation functions}

\author{R. W. Robinett and L. C. Bassett}
%\email{rick@phys.psu.edu}
\affiliation{
Department of Physics\\
The Pennsylvania State University\\
University Park, PA 16802 USA \\
}

\date{\today}

\begin{abstract}
The autocorrelation function, $A(t)$, measures the overlap (in Hilbert space)
of a time-dependent quantum mechanical wave function, $\psi(x,t)$, 
with  its initial value, $\psi(x,0)$. It finds extensive use in the 
theoretical analysis and experimental measurement of such phenomena
as quantum wave packet revivals. We evaluate explicit expressions for the 
autocorrelation function for time-dependent Gaussian solutions of the 
Schr\"{o}dinger equation corresponding to the cases of a free particle, 
a particle undergoing uniform acceleration, a particle in a harmonic 
oscillator potential, and a system corresponding to an unstable equilibrium
(the so-called `inverted' oscillator.)
We emphasize the importance of momentum-space methods where such 
calculations are often more straightforwardly realized, as well as stressing
their role in providing complementary  information to results 
obtained using position-space wavefunctions.

\end{abstract}

\maketitle

\section{\label{sec_intro} Introduction}

Despite widely varying approaches to the subject, every introductory
quantum mechanics text of which we are aware discusses the subject of
wave function normalization and conservation of probability, proving
that
\begin{equation}
\langle \psi_t | \psi_t \rangle
= 
\langle \psi_0 | \psi_0 \rangle
= 1
\end{equation}
provided the Hamiltonian describing the system is Hermitian. This
important relationship is perhaps most often discussed in position-space
using
\begin{equation}
\langle \psi_t|\psi_t \rangle
= 
\int_{-\infty}^{+\infty}\, \psi^{*}(x,t) \, \psi(x,t)\,dx
= 1
\end{equation}
but,  using the Fourier transform, 
\begin{equation}
\psi(x,t) = \pre \integral e^{ipx/\hbar} \, \phi(p,t)\,dp
\, ,
\label{fourier_transform}
\end{equation}
it is equally well written in the form
\begin{equation}
\langle \psi_t|\psi_t \rangle
= 
\int_{-\infty}^{+\infty}\, \phi^{*}(p,t) \, \phi(p,t)\,dp
= 1
\, .
\end{equation}
For bound state problems, where any time-dependent state can be expanded
in terms of energy eigenstates, $u_{n}(x)$, with quantized energies, $E_n$,
we can also write
\begin{equation}
\psi(x,t) = \sum_{n} a_n \, u_n(x)\, e^{-iE_n t/\hbar}
\qquad
\mbox{where}
\qquad
a_{n} = \int_{-\infty}^{+\infty} [u_{n}(x)]^*\, \psi(x,0) \, dx
\label{expansion_in_eigenstates}
\end{equation}
so that the normalization condition is given by
\begin{equation}
\langle \psi_t | \psi_t \rangle = \sum_{n=1}^{\infty}|a_n|^2
= 1
\,.
\end{equation}

An important related concept, namely the autocorrelation function
\cite{nauenberg}, $A(t)$, is defined as the overlap of a 
time-dependent
state, $\psi_t$, with its initial value, $\psi_{0}$, namely
\begin{equation}
A(t) \equiv \langle \psi_t |\psi_0 \rangle
= 
\int_{-\infty}^{+\infty}\, \psi^{*}(x,t) \, \psi(x,0)\,dx
=
\int_{-\infty}^{+\infty}\, \phi^{*}(p,t) \, \phi(p,0)\,dp
\end{equation}
and is extensively used in the study of the time-development of
quantum wave packets in bound state systems. One clearly has
$A(0) = 1$, by definition, if the initial wave function is itself
properly normalized, but generally $|A(t)| < 1$ for later times as
the wave packet develops in time  and the different energy/momentum 
components evolve differently.
Besides being of obvious theoretical value in the analysis of 
time-dependent systems, the autocorrelation function is physically
important because it is very directly related to the observable
ionization signal in the pump-probe type experiments where the
time-development of atomic wave packets is studied experimentally 
\cite{alber_original}, \cite{yeazell_detection}.

For a single (bound) energy eigenstate or stationary state, 
where $\psi(x,t) = u_n(x)e^{-iE_nt/\hbar}$,
one clearly has 
\begin{equation}
A(t) = e^{+iE_nt/\hbar}
\qquad
\mbox{giving}
\qquad
|A(t)|^2 = 1
\, . 
\end{equation}
For more general bound state systems as in 
Eqn.~(\ref{expansion_in_eigenstates}), however, 
the most appropriate form is
\begin{equation}
A(t) = \langle \psi_t |\psi_0 \rangle
=
\sum_{n=1}^{\infty}|a_n|^2 \, e^{+iE_n t/\hbar}
\end{equation}
so that information on the non-trivial time-development of the 
wave packet is encoded in the energy eigenvalue spectrum. 
For wave packets which are constructed from
energy eigenstates centered around some large value of $n = n_0$, one can
write
\begin{equation}
E(n) \approx E(n_0) + E'(n_0)(n-n_0) + \frac{E''(n_0)}{2}(n-n_0)^2
+ \cdots
\label{1d_periods}
\end{equation}
which gives the time-dependence of each individual quantum eigenstate as
\begin{eqnarray}
e^{-iE_nt/\hbar} & = & \exp\left( -i/\hbar\left[E(n_0)t + (n-n_0) E'(n_0)t +
\frac{1}{2} (n-n_0)^2 E''(n_0) t + \cdots \right]\right) \nonumber \\
& = & \exp \left( -i\Omega_0 t - 2\pi i(n-n_0) t/T_{cl}
- 2\pi i(n-n_0)^2t/T_{rev} + \cdots\right)
\end{eqnarray}
in terms of which the classical period and quantum mechanical revival
time (discussed below) are given respectively by
\begin{equation}
T_{cl} = \frac{2\pi\hbar}{|E'(n_0)|}
\qquad
\mbox{and}
\qquad
T_{rev} = \frac{2\pi \hbar}{|E''(n_0)|/2}
\label{various_periods}
\end{equation}
and the common first term, $\exp(-i\Omega_0 t) = \exp(-iE(n_0)t/\hbar)$,
is an unimportant overall phase. The most familiar example is the
harmonic oscillator, with $E_n = (n+1/2)\hbar \omega$, 
and any wavepacket in this system is periodic 
with period $T_{cl} = 2\pi/\omega$
and $T_{rev} \rightarrow \infty$ since $E''(n_0) = 0$. 

This form of the autocorrelation function is especially useful in the 
context of quantum wave packet revivals \cite{nauenberg}, that is, systems 
where initially localized states which have a short-term, quasi-classical 
time evolution, can spread significantly over several orbits, only to 
reform later in the form of a quantum revival in which the spreading 
reverses itself, the wave 
packet relocalizes, and the semi-classical periodicity is once again evident.
The presence of an approximate quantum revival at a later time, $t = T_{rev}$,
is indicated by $|A(T_{rev})| \approx 1$, accompanied by evidence of the
return of the short-term periodicity in $A(t)$ with  the classical period,
$T_{cl}$. The phase-structure of $A(t)$ can also yield  useful information,
as parametric plots of $Re[A(t)]$ versus $Im[A(t)]$ (Argand diagrams) can
provide  striking visualizations \cite{physics_report_revival}
of the highly correlated Schr\"odinger
cat-like states which evolve at fractional multiples of the
revival time, the mathematics of which was first worked out in detail
by Averbukh and Perelman \cite{averbukh}.

Such revival phenomena have been observed in a wide variety of physical
systems, especially in Rydberg atoms \cite{yeazell} --
\cite{wals}. In many cases,
the autocorrelation function is closely related to experimentally observable
quantities \cite{alber_original}, \cite{yeazell_detection}, 
and its use has become very familiar for analyzing model systems
exhibiting exact or approximate wave packet revivals.

We note that early investigators commented on the existence of a `quantum
recurrence theorem' \cite{recurrence} which utilized the notion of
`distance' (in Hilbert space) between a time-dependent quantum state and 
its initial value as 
\begin{eqnarray}
||\psi_{t} - \psi_{0}||^2 & = & \int_{-\infty}^{+\infty} |\psi(x,t) - \psi(x,0)|^2\,dx  \nonumber \\
& = & 
\int_{-\infty}^{+\infty} |\psi(x,t)|^2\,dx 
+
\int_{-\infty}^{+\infty} |\psi(x,0)|^2\,dx \nonumber \\ 
& & \,\,\,\,\,\,\,\,
- 
\int_{-\infty}^{+\infty} \psi^{*}(x,t)\psi(x,0)\,dx
-
\int_{-\infty}^{+\infty} \psi^{*}(x,0)\psi(x,t)\,dx \\ 
& = & 
2\left(
1-\Re\left[\int_{-\infty}^{+\infty} \psi^*(x,t)\,\psi(x,0)\,dx\right]\right)
= 2(1-\Re[A(t)])
\nonumber 
\end{eqnarray}
which is also related to $A(t)$; we also note that Baltz \cite{baltz} has 
considered similar ideas in a more pedagogical context. 

In contrast to the research literature, where $A(t)$ is now a standard
tool, there are few, if any, examples of the evaluation of the autocorrelation
function for the many familiar and frequently studied  model systems of 
introductory quantum mechanics. Besides being useful as a diagnostic
for the rate of time-evolution of a quantum state, such calculations 
of $A(t)$ can also help answer questions
such as {\it How similar are two quantum states?}, both in magnitude and
phase, and we will stress the complementary roles that position-space and
momentum-space approaches can have in addressing such issues in what
follows.

In this note, we will use the special properties of Gaussian wave packet 
solutions to explicitly evaluate $A(t)$, obtaining closed form expressions,
for four familiar and accessible model systems. We focus on the cases of a 
free particle 
(Sec.~II), 
a particle undergoing uniform acceleration (Sec.~III), 
a particle in a harmonic oscillator potential (Sec.~IV), 
and a system corresponding to an unstable equilibrium (the `inverted'
oscillator, in Sec.~V.) The first two cases do not correspond to
 bound state systems, but do provide useful results for comparison to
the short-term time-dependence of wave packets in systems such as the
infinite well \cite{physics_report_revival}, 
\cite{segre} -- \cite{different_fractional}
and the so-called `quantum bouncer' \cite{chen_gravity} --
\cite{robinett_bouncer}
where, for small times at least, the wave packet propagation is similar
to the corresponding unbound case. For the case of the harmonic oscillator,
any wave packet solution (Gaussian or not) is explicitly periodic with
$T_{cl} = 2\pi/\omega$ and the wavepacket never enters a truly `collapsed'
phase; the oscillator does, however, provide a useful explicit example 
illustrating the exactly periodic behavior of $A(t)$,
seen more approximately in many bound state systems.
For example, Nauenberg \cite{nauenberg}
has provided an elegant general description of the medium-term 
time-development of $A(t)$ for a general, one-dimensional, bound state system
and the exact results presented here for the harmonic oscillator 
can be used as an efficient `benchmark' for comparison to that more
general analysis.
While we will focus on closed-form results for oscillator wave packets
(obtained using propagator methods), we can also make contact with the 
expansion in eigenstates in Eqn.~(\ref{expansion_in_eigenstates}) as well.

Such studies of the general behavior of $A(t)$ for many standard
examples are also useful as they complement existing work on 
the rate of wave packet spreading \cite{asy_1} -- \cite{asy_3}
and especially on the time evolution of quantum states
\cite{rate_1}, \cite{rate_2}. In this last context, the example
provided here for the free particle (in Sec.~II) can be used as
a specific case to confirm a (hopefully) well-known result for isolated
quantum systems \cite{mandelstam} -- \cite{kamal}, namely
\begin{equation}
|\langle \psi_{t}|\psi_{0}\rangle|^2
 \geq
\cos^2\left(\frac{\Delta H t}{\hbar}\right)
\qquad
\mbox{for}
\qquad
0 \leq t \leq \frac{\pi \hbar}{2\Delta H}
\label{mandelstam}
\end{equation}
where $\Delta H = \sqrt{\langle H^2\rangle - \langle H\rangle^2}$
is the uncertainty in the free-particle energy of the wave packet.

Because our presentation here will use the same notation and many of the
same methods as the companion paper \cite{bassett}, we will refer 
extensively to results from that paper, especially for properties of the 
standard Gaussian wave packets we utilize.

\section{\label{sec_free_particle} Free-particle Gaussian wave packets}

The most general free-particle, momentum-space and position-space Gaussian
wave packets, with arbitrary initial values of $\langle x \rangle_0 = x_0$
and $\langle p \rangle_0 = p_0$, can be written \cite{bassett} in the form 
\begin{eqnarray}
\phi(p,t) = \phi_0(p)e^{-ip^2t/2m\hbar} 
& = &  
\sqrt{\frac{\alpha}{\sqrt{\pi}}}
\, e^{-\alpha^2(p-p_0)^2/2}
\, e^{-ipx_0/\hbar}
\, e^{-ip^2t/2m\hbar} 
\label{p_gaussian_t} \\ 
\phi(p,0) & = & 
\sqrt{\frac{\alpha}{\sqrt{\pi}}}
\, e^{-\alpha^2(p-p_0)^2/2}
\, e^{-ipx_0/\hbar}
\label{p_gaussian_0}
\end{eqnarray}
and
\begin{eqnarray}
\psi(x,t) & = &  \frac{1}{\sqrt{\sqrt{\pi} \alpha \hbar (1+it/t_0)}}
\,
e^{ip_0(x-x_0)/\hbar}
\, e^{-ip_0^2t/2m\hbar}
\,
e^{-(x-x_0-p_{0}t/m)^2/2(\alpha \hbar)^2(1+it/t_0)}
\label{x_gaussian_t} \\
\psi(x,0) & = &  \frac{1}{\sqrt{\sqrt{\pi} \alpha \hbar}}
\,
e^{ip_0(x-x_0)/\hbar}
\,
e^{-(x-x_0)^2/2(\alpha \hbar)^2}
\label{x_gaussian_0}
\end{eqnarray}
where $t_0 \equiv m\hbar \alpha^2$ defines the spreading time.
The calculation of $A(t)$ is done most straightforwardly in
momentum-space where
\begin{eqnarray}
A(t) & = &  \int_{-\infty}^{+\infty}\, \phi^{*}(p,t) \, \phi(p,0)\,dp
= \frac{\alpha}{\sqrt{\pi}}\int_{-\infty}^{+\infty} \, 
e^{-\alpha^2(p-p_0)^2} \, e^{ip^2t/2m\hbar} \nonumber \\ 
& = &  \frac{1}{\sqrt{1-it/2t_0}} 
\exp\left[
\frac{i\alpha^2 p_0^2t}{2t_0(1-it/2t_0)}
\right]
\label{free_particle_momentum}
\end{eqnarray}
and the modulus-squared is then given by
\begin{equation}
|A(t)|^2 = \frac{1}{\sqrt{1+(t/2t_0)^2}}
\exp\left[-2\alpha^2 p_0^2 \frac{(t/2t_0)^2}{(1+(t/2t_0)^2)}\right]
\, .
\label{free_autocorrelation_modulus}
\end{equation}
The same result can, of course, be obtained in position-space, with
\begin{equation}
A(t) = \int_{-\infty}^{+\infty} \psi^{*}(x,t)\,\psi(x,0)\,dx =
\frac{1}{\sqrt{1-it/2t_0}}
\,
e^{ip_0^2 t/2m\hbar}
\, 
\exp\left[\frac{-(p_0 t/m)^2}{4(\alpha \hbar)^2 (1-it/2t_0)}\right]
\label{free_autocorrelation_function}
\end{equation}
which is easily seen to be identical to Eqn.~(\ref{free_particle_momentum})
with a minimum of manipulation.

We note that for times satisfying $ t << t_0$  there is an 
increasing exponential suppression of the overlap between $\psi_t$
and $\psi_0$, but the exponential factor  does `saturate' for
long times, giving
\begin{equation}
|A(t>>2t_0)|^2 \longrightarrow \frac{2t_0}{t} 
\exp\left[-\frac{p_{0}^2}{\Delta p_0^2}\right]
\quad
\mbox{since}
\quad
\Delta p_0 = \frac{1}{\alpha \sqrt{2}}
\, . 
\end{equation}
The asymptotic form of the exponential factor can be understood 
by noting that the `distance in position space'  between the initial
`peak' at $\langle x \rangle_0 = x_0$, 
and that at later times when $\langle x \rangle_t = x_0 + p_0t/m$,  grows
linearly with $t$, while for long times the position spread,
\begin{equation}
\Delta x_t = \Delta x_0 \sqrt{ 1 + (t/t_0)^2}
\quad
\longrightarrow
\quad
\Delta x_0 \frac{t}{t_0}
\, , 
\end{equation}
grows in the same way. This leads to factors in the exponent of the
form
\begin{equation}
\frac{(x(t) - x(0))^2}{(\Delta x_t)^2} 
\longrightarrow
\frac{(p_0t/m)^2}{(\Delta x_0 (t/t_0))^2}
\approx
\left(\frac{p_0 t_0}{m \Delta x_0}\right)^2
\approx (p_0 \alpha)^2
\quad
\mbox{since}
\quad
\Delta x_0 = \frac{\alpha \hbar}{\sqrt{2}}
\, .
\end{equation}

This argument is most transparent using the position-space wave
functions where the exponentially small overlap of {\it magnitudes} is
clear, but the same suppression arises in the momentum-space formulation,
this time due to cancellations arising from the rapidly oscillating 
{\it phase} factor in the integrand in Eqn.~(\ref{free_particle_momentum}). 
We will occasionally distinguish the exponential suppression factors 
(which we can describe as `dynamical'
as they depend on the initial wave packet parameter $p_0$) from the more
intrinsic pre-factor term (containing only the spreading time) which is
due to the natural dispersion of the wave packet (which we can therefore
describe as `dispersive'.)

For this case of an isolated quantum system, the specific result for the
autocorrelation function in Eqn.~(\ref{free_autocorrelation_modulus})
must also satisfy the general theorem in Eqn.~(\ref{mandelstam}). To
lowest non-trivial order in $t$ (${\cal O}(t^2)$), the modulus of $A(t)$ 
in this case for short times is
\begin{eqnarray}
|A(t)|^2 & =&  \frac{1}{\sqrt{1+(t/2t_0)^2}}
\exp\left[-2\alpha^2 p_0^2 \frac{(t/2t_0)^2}{(1+(t/2t_0)^2)}\right] 
\nonumber \\
& \longrightarrow & 
\left(1 - \frac{2\alpha^2 p_0^2 t^2}{4t_0^2} + \cdots \right)
\left( 1 - \frac{t^2}{8t_0^2} + \cdots\right) 
\label{explicit} \\
& \approx & 
 1 - 
\frac{\alpha^2 t^2}{2t_0^2} \left(p_0^2 + \frac{1}{4\alpha^2}
\right)
+\cdots 
\, .
 \nonumber
\end{eqnarray}
We note that this result arises from both the exponential (`dynamical') 
suppression, as well as the (`dispersive') prefactor term.
The general result of Eqn.~(\ref{mandelstam}) requires the calculation of
\begin{equation}
\langle H \rangle  = \frac{1}{2m} \left( p_0^2 + \frac{1}{2\alpha^2}\right)
\qquad
\quad
\mbox{and}
\qquad
\quad
\langle H^2 \rangle  =  \left(\frac{1}{2m}\right)^2
\left(  p_0^4 + \frac{3p_0^2}{\alpha^2} + \frac{3}{4\alpha^4}\right) 
\end{equation}
which give
\begin{equation}
(\Delta H)^2  =  \left(\frac{1}{2m}\right)^2\frac{2}{\alpha^2}
\left( p_0^2 + \frac{1}{4\alpha^2}\right)
\, . 
\end{equation}
When the right-hand-side of Eqn.~(\ref{mandelstam}) is expanded to
lowest order in $t$ (again, ${\cal O}(t^2)$), using this result, 
it is found that 
\begin{eqnarray}
\cos^2\left(\frac{\Delta H t}{\hbar}\right)
& = &
\left( 
1  - \frac{1}{2} \left( \frac{\Delta H t}{\hbar}\right)^2
+ \cdots \right)^2 \nonumber \\
& = &
1 - \frac{1}{2 m^2  \alpha^2}\left( p_0^2 + \frac{1}{4\alpha^2}
\right)
\left(\frac{t^2}{\hbar^2}\right)
+ \cdots \\
& = &
1 - \frac{\alpha^2t^2}{2 t_0^2}\left( p_0^2 + \frac{1}{4\alpha^2}
\right) 
+ \cdots 
\nonumber 
\end{eqnarray}
since $t_0 \equiv m \hbar \alpha^2$ and the
result in Eqn.~(\ref{explicit}) is found to satisfy (in fact, to saturate,
at this order) the inequality in Eqn.~(\ref{mandelstam}) for short times, 
confirming the general result. We note that it is easy to show that the
inequality is also satisfied to ${\cal O}(t^4)$ where the left-hand-side is
indeed larger than the right at this order.

\section{\label{sec_uniform_acceleration} Uniform acceleration}

The explicit form for Gaussian solutions, in both
momentum-space and position-space, for the problem of a particle
undergoing uniform acceleration (constant force $F$ or linear potential
given by $V(x) = -Fx$) are given in Ref.~\cite{bassett}. For the
momentum-space form we have
\begin{eqnarray}
\phi(p,t) = \Phi(p-Ft) e^{-ip^3t/6mF\hbar} 
& = &  
\sqrt{\frac{\alpha}{\sqrt{\pi}}}
\, e^{-\alpha^2((p-Ft) - p_0)^2/2}
\, e^{-i(p-Ft)x_0/\hbar}
\, e^{i((p-Ft)^3 - p^3)/6mF\hbar}
\label{p_acceleration_t} \\
\phi(p,0) & = & 
\sqrt{\frac{\alpha}{\sqrt{\pi}}}
\, e^{-\alpha^2(p- p_0)^2/2}
\, e^{-ipx_0/\hbar}
\label{p_acceleration_0} 
\end{eqnarray}
with arbitrary initial position ($x_0$) and momentum ($p_0$).
The corresponding position-space wavefunction is
\begin{eqnarray}
\psi(x,t) & = & \left[e^{iFt(x_0-Ft^2/6m)/\hbar}\, e^{i(p_0+Ft)(x-x_0 - p_0t/2m)/\hbar}\right] \left(
\frac{1}{\sqrt{\sqrt{\pi}\alpha \hbar (1+it/t_0)}}
\right) \nonumber \\ 
& & \,\,\,
\times
\, e^{-(x-(x_0+p_0t/m+Ft^2/2m))^2/2(\alpha \hbar^2)^2(1+it/t_0)}
\, . 
\label{x_acceleration_t}
\end{eqnarray}
with $\psi(x,0)$ given in Eqn.~(\ref{x_gaussian_0}).

The calculation of $A(t)$ can be done using either form to obtain
\begin{equation}
A(t)  =  \frac{1}{\sqrt{1-it/2t_0}} 
\exp\left[\frac{(2ip_0^2t/m\hbar - (\alpha F t)^2 (1+(t/2t_0)^2))
}{4(1-it/2t_0)} 
\right] 
\, 
e^{-iFt(x_0-Ft^2/6m)/\hbar} 
\, , 
\end{equation}
and the same factors of $1-it/2t_0$ as in 
Eqn.~(\ref{free_particle_momentum})
are obtained; this expression also
reduces to that case in the free particle
limit  when $F \rightarrow 0$,  as it must. The modulus-squared is given by 
\begin{equation}
|A(t)|^2 = \frac{1}{\sqrt{1+(t/2t_0)^2}}
\,
\exp\left[-2\alpha^2 (p_0^2 +(Ft_0)^2(1+(t/2t_0)^2)
\left(\frac{(t/2t_0)^2}{1+(t/2t_0)^2}\right) 
\right]
\label{acceleration_autocorrelation_function}
\end{equation}
and we note that this result can be obtained from 
Eqn.~(\ref{free_autocorrelation_modulus}) by the simple substitution
\begin{equation}
p_0^2 \longrightarrow
p_0^2 +(Ft_0)^2(1+(t/2t_0)^2)
\, . 
\end{equation}
For this case of uniform acceleration, the wave packet spreading is
identical (same $\Delta x_t$) as in 
the free-particle case, which can be  understood
by noting that the distance between two classical particles starting at the 
same initial location, undergoing the same force, but with slightly different
initial velocities (or momenta, $p_0^{(A)} - p_0^{(B)} =
\Delta p_0$) would be
\begin{equation}
x_A(t) - x_B(t) = (x_0 + p_0^{(A)}t/m + Ft^2/2m) 
                - (x_0 + p_0^{(B)}t/m + Ft^2/2m) 
= \frac{\Delta p_0t}{m}
\end{equation}
which increases linearly with time, in exactly the same way
as for the  free-particle solutions (when $F=0$).
The `distance' between the peaks in $\psi_0$ and $\psi_t$, however,
eventually grows as $t^2$ so that the exponential (`dynamical') 
suppression in $A(t)$ does not saturate, while the `dispersive' 
pre-factor is the same as for the free-particle case. 

We also note that the factors of $p_0$ and $F$ in $|A(t)|^2$  
appear in quadrature,
and not in a combination such as $p_0+Ft$. One might naively expect that
in cases where $p_0$ and $F$ have opposite signs, so that at a time
given by $t_{ret} = 2p_0/F$ when the classical particle 
(and central value of the
quantum wave packet) has returned to $x_0$, the magnitudes of 
$\psi(x,t)$ and $\psi(x,0)$ would be similar (only differing in possibly
small spreading effects due to the $(1+it/t_0)$ factors) and so would give
rise to a relatively large value of $|A(t)|$. While this does indeed
occur for the {\it magnitudes}, at that time  the
classical momentum is of the opposite sign ($p(0) = p_0 \rightarrow
p(t_{ret}) = -p_0$),  giving rise to rapidly oscillating phase factors 
in the quantum wave function, which still gives the expected exponential 
suppression.

\section{\label{sec_sho} Simple harmonic oscillator wave packets}

The free particle and case of uniform acceleration, corresponding to 
unbound motions, do not provide examples for direct comparison to 
the (quasi) periodic behavior seen in bound state systems. The harmonic 
oscillator admits Gaussian wave packet solutions which can be written 
(using propagator techniques) in closed form for arbitrary initial values 
of position and momentum ($x_0$, $p_0$) and for 
which the evaluation of $A(t)$ is therefore possible. In this case, 
we can also
write any general wave packet as an expansion in eigenstates as in
Eqn.~(\ref{expansion_in_eigenstates}), using $E_n = (n+1/2)\hbar \omega$,
as 
\begin{equation}
\psi(x,t) = \sum_{n=0}^{\infty} a_n u_n(x) e^{-i(n+1/2)\omega t}
\label{sho_expansion}
\end{equation}
from which it is clear that the observable probability density,
$|\psi(x,t)|^2$, is periodic with the classical period,
$T_{cl} = 2\pi/\omega$. 

Using the propagator techniques outlined in \cite{bassett}, and the 
initial position-space wave function
\begin{equation}
\psi(x,0) = \frac{1}{\sqrt{\beta \sqrt{\pi}}}
\, e^{ip_0x/\hbar}
\,
e^{-(x-x_0)^2/2\beta^2}
\, ,
\end{equation}
where $\beta \equiv \alpha \hbar$, one can evaluate the time-development
in closed form as 
\begin{equation}
\psi(x,t) = \frac{1}{\sqrt{L(t)\sqrt{\pi}}}
\exp\left[
\frac{S[x,t]}{2\beta L(t)}
\right]
\label{general_sho_solution}
\end{equation}
where
\begin{equation}
L(t) \equiv \beta \cos(\omega t)
+ \frac{i \hbar}{m\omega \beta} \sin(\omega t)
\end{equation}
and
\begin{eqnarray}
S[x,t] & \equiv &
-x_0^2\cos(\omega t)
+2xx_0
- x^2\left[\cos(\omega t) +\frac{im\omega \beta^2\sin(\omega t)}{\hbar}\right] \\
& &
\,\,\,\,\,\,\,\,\,\,\,\,\,
- \frac{ 2x_0p_0 \sin(\omega t)}{m \omega}
+ \frac{2i\beta^2 p_0x}{\hbar}
- \frac{i\beta^2 p_0^2 \sin(\omega t)}{m\omega \hbar} 
\nonumber 
\, .
\end{eqnarray}
The corresponding position-space probability density can be written as 
\begin{equation}
|\psi(x,t)|^2
= \frac{1}{\sqrt{\pi} |L(t)|}
\exp\left[
-\frac{(x-x_0\cos(\omega t) - p_0 \sin(\omega t)/m\omega)^2}{|L(t)|^2}
\right]
\end{equation}
with
\begin{equation}
\langle x \rangle_t = x_0 \cos(\omega t) + \frac{p_0 \sin(\omega t)}{m\omega}
\qquad
\quad
\mbox{and}
\qquad
\quad
\Delta x_t = \frac{|L(t)|}{\sqrt{2}}
\,.
\end{equation}
Thus, the expectation value moves in accordance with classical
expectations \cite{styer_2}, while the width oscillates 
(from wide to narrow, or 
vice versa.) For the special case of the `minimum uncertainty' wave packet
where 
\begin{equation}
\beta^2 = \frac{\hbar}{m\omega} \equiv \beta_0^2
\, , 
\label{minimum_uncertainty}
\end{equation}
the width of the packet is fixed as
\begin{equation}
\Delta x_t = \Delta x_0 = \frac{\beta_0}{\sqrt{2}}
\end{equation}
which is the same as the ground state oscillator energy eigenvalue state,
but simply oscillates at the classical frequency. 
We note that the momentum-space wavefunctions can also be written
using a propagator formalism \cite{robinett} and used to evaluate
$A(t)$ in a parallel fashion, obtaining the same result.

The evaluation of $|A(t)|^2$ for general values of $\beta$, $x_0$, and
$p_0$ is straightforward enough, but the resulting expressions 
are somewhat cumbersome, so we will focus on several special cases
as illustrative.

\vskip 0.8cm
\noindent
{\bf Case I}: Minimum uncertainty wave packets, $\beta = \beta_0$.

In this case the evaluation of $A(t)$ gives
\begin{equation}
A(t) = \sqrt{\cos(\omega t) + i \sin(\omega t)}
\exp\left[
- \left(\frac{x_0^2}{2\beta_0^2} + \frac{\beta_0^2p_0^2}{2\hbar^2}\right)
[(1-\cos(\omega t)) - i\sin(\omega t)]
\right]
\end{equation}
where great simplifications have been made by noting that
\begin{equation}
\frac{1}{\cos(\omega t) -i \sin(\omega t)}
= \cos(\omega t) + i \sin(\omega t)
\, . 
\end{equation}
We note that a very similar expression arises in analyses of the
macroscopic wavefunction for Bose-Einstein condensates \cite{bec_2}
and the collapse and revival of the matter wave field for such systems
has been observed experimentally \cite{bec_revivals}.

Once again, the two important parameters appear together in quadrature,
as in the uniform acceleration case. This gives
\begin{equation}
|A(t)|^2
= \exp
\left[
-\left(
\frac{x_0^2}{\beta_0^2} + \frac{\beta_0^2p_0^2}{\hbar^2}\right)
[1-\cos(\omega t)]\right]
\label{sho_case}
\end{equation}
which clearly exhibits the expected periodicity. All of the suppression
can be attributed to the `dynamical' factors (those in the exponential,
containing $x_0$ and $p_0$) as there is no 'dispersive' pre-factor
component for this constant width packet.

For this case, the
minimum degree of overlap at any point during a single classical
period is 
\begin{equation}
|A(T_{cl}/2)|^2 
= 
\exp
\left[
-2\left(\frac{x_0^2}{\beta_0^2} + \frac{\beta_0^2 p_0^2}{\hbar^2}\right)
\right]
\end{equation}
so there is no time at which the wave packet is ever truly orthogonal
to its initial state.

\vskip 0.8cm

\noindent
{\bf Case II}: Arbitrary $\beta$, but $x_0, p_0 = 0$. For this case, the
wave packet does not oscillate, but only `pulsates', and 
the time-dependent wave function simplifies to 
\begin{equation}
\psi(x,t) = 
\frac{1}
{\sqrt{\pi}\sqrt{(\beta \cos(\omega t) +(i\hbar/m \omega \beta)\sin(\omega t))}}
\, 
\exp
\left[\frac{-x^2 [\cos(\omega t) +(im\omega \beta^2/\hbar)\sin(\omega t)]}
{2\beta^2 [\cos(\omega t) +(i\hbar/m \omega \beta)\sin(\omega t)]}
\right] 
\end{equation}
It is convenient to define the parameters
\begin{equation}
r \equiv \frac{\hbar}{m \omega \beta^2} = \frac{ \beta_0^2}{\beta^2}
\qquad
\mbox{so that}
\qquad
\frac{1}{r} = \frac{\beta^2}{\beta_0^2}
\end{equation}
in terms of which the resulting autocorrelation function in this case
has the very simple form
\begin{equation}
A(t) = \sqrt{\frac{2}{2\cos(\omega t) -i(r+1/r)\sin(\omega t)}}
\label{special_case}
\end{equation}
or
\begin{equation}
|A(t)|^2 = \frac{1}{\sqrt{\cos^2(\omega t) + (r+1/r)^2 \sin^2(\omega t)/4}}
\end{equation}
all of which can be attributed to a `dispersive' (but in this case
periodic) pre-factor.

We first note that in this case $A(t)$ is invariant under the transformation
$ r \rightarrow 1/r$, in other words, the time-dependence is the same
for both initially wide ($\beta > \beta_0$) or narrow ($\beta < \beta_0$)
packets. It is clear that the larger the deviation from the 
`minimum uncertainty'
wavepacket, the faster the wavepacket `pulsates' away from its initial
shape. It is also noteworthy that in this case $|A(T_{cl}/2)|=1$ so that
the wave packet returns to its initial form (up to a constant complex
phase) {\it twice} each classical period. This can be understood from the 
expansion of this wave form in terms of energy eigenstates. In this case,
where the parameters $x_0,p_0$ both vanish, one is expanding an even-parity
function in Eqn.~(\ref{sho_expansion}),  
so that only the even ($a_{2n}$) terms are nonvanishing and 
the $n$-dependent exponential factors in 
Eqn.~(\ref{expansion_in_eigenstates}) oscillate  twice as rapidly
as in the general case. 

Finally, for the very special case  where $\beta = \beta_0$ ($r=1$) as well, 
we recover the ground state energy eigenstate of the oscillator, with its 
trivial stationary-state time-dependence
($\psi_0(x,t) = u_0(x)\exp(-iE_0t/\hbar)$) and 
Eqn.~(\ref{special_case}) indeed reduces to
\begin{equation}
A(t) \stackrel{r\rightarrow 1}{\longrightarrow}
\sqrt{\frac{2}{2\cos(\omega t) - 2i\sin(\omega t)}}
= \sqrt{e^{i\omega t}} = e^{+i\omega t/2}
\end{equation}
as expected. 

Wave packet solutions (Gaussian or not) of the harmonic oscillator
can be shown (using the expansion in Eqn.~(\ref{sho_expansion}), for
example) to satisfy \cite{saxon}
\begin{equation}
\psi(x,t+mT_{cl}) = (-1)^{m}\, \psi(x,t)
\label{sho_standard}
\end{equation}
with a similar result for the momentum-space version as well. Because
of the specially symmetric nature of the potential, we also have
\begin{equation}
\psi(-x,t+T_{cl}/2) = (-i) \, \psi(x,t)
\qquad
\mbox{and}
\qquad
\phi(-p,t+T_{cl}/2) = (-i) \, \phi(p,t)
\label{sho_anticorrelation}
\end{equation}
so that half a period later, the wave-packet is reproduced, but at
the opposite `corner' of phase space, namely with $x \leftrightarrow
-x$ and $p \leftrightarrow -p$: we note that two applications of
Eqn.~(\ref{sho_anticorrelation}) reproduce Eqn.~(\ref{sho_standard}).
One can also show these connections using the propagator techniques
in Ref.~\cite{bassett}, provided one properly identifies the
complex pre-factors as described in detail in Ref.~\cite{saxon}.

This type of behavior can be diagnosed using a variation on the
standard autocorrelation function, namely 
\begin{equation}
\overline{A}(t) 
\equiv
\int_{-\infty}^{+\infty}\, \psi^{*}(-x,t) \, \psi(x,0)\,dx
=
\int_{-\infty}^{+\infty}\, \phi^{*}(-p,t) \, \phi(p,0)\,dp
\end{equation}
which measures the overlap of the initial state with the `out-of-phase'
version of itself at later times. Given the simple connections in
Eqn.~(\ref{sho_anticorrelation}), we can immediately write, for 
Case I considered above, 
\begin{equation}
|\overline{A}(t)|^2 
= \exp
\left[
-\left(
\frac{x_0^2}{\beta_0^2} + \frac{\beta_0^2 p_0^2}{\hbar^2}\right)
(1+\cos(\omega t)\right]
\end{equation}
which is exponentially suppressed at integral multiples of $T_{cl}$,
but unity at $t = (2k+1)T_{cl}/2$. This type of {\it anti-correlation}
function finds use in the study of wave packet revivals 
\cite{physics_report_revival} where quantum
wave packets may reform near $t = T_{rev}$, as 
in Eqn.~(\ref{various_periods}), but out of phase with the 
original packet.

\section{\label{sec_unstable} `Inverted' oscillator wave packets for 
unstable equilibrium}

As described in detail in Ref.~\cite{bassett}, the case of the `inverted'
oscillator, defined by
\begin{equation}
\tilde{V}(x) \equiv  - \frac{1}{2} m\tilde{\omega}^2 x^2
\end{equation}
can be studied using wave packet results for the standard harmonic oscillator
by making the substitutions
\begin{equation}
\omega \rightarrow i \tilde{\omega}
\, ,
\qquad
\sin(\omega t) \rightarrow i \sinh(\tilde{\omega} t)
\, ,
\qquad
\mbox{and}
\qquad
\cos(\omega t) \rightarrow \cosh(\tilde{\omega} t)
\, . 
\end{equation}
The general wave packet solution in Eqn.~(\ref{general_sho_solution}), for
example, can be carried over in this way to obtain the `runaway' wavepacket,
with probability density given by
\begin{equation}
|\psi(x,t)|^2
= \frac{1}{\sqrt{\pi} |B(t)|}
\exp\left[
-\frac{(x-x_0\cosh(\tilde{\omega} t) 
- p_0 \sinh(\tilde{\omega} t)/m\tilde{\omega})^2}{|B(t)|^2}
\right]
\end{equation}
with
\begin{equation}
\langle x \rangle_t = x_0 \cosh(\tilde{\omega} t) + 
\frac{p_0 \sinh(\tilde{\omega} t)}{m\tilde{\omega}}
\qquad
\quad
\mbox{and}
\qquad
\quad
\Delta x_t = \frac{|B(t)|}{\sqrt{2}}
\end{equation}
where
\begin{equation}
|B(t)| = \sqrt{\beta^2 \cosh^2(\tilde{\omega} t) 
+ (\hbar/m\tilde{\omega} \beta)^2
\sinh^2(\tilde{\omega} t)}
\,.
\end{equation}
As above, the expression for $A(t)$ for the general case is cumbersome,
so we only examine it for one specific case as an example, namely the
case where $\beta = \beta_0 = \sqrt{h/m\tilde{\omega}}$. 
This situation no longer corresponds to a constant width wave packet, since
\begin{equation}
\Delta x_t \longrightarrow \frac{\beta_0}{\sqrt{2}}
\sqrt{\cosh^2(\tilde{\omega} t) + \sinh^2(\tilde{\omega} t)}
\end{equation}
increases exponentially, as the individual momentum 
components comprising the wave packet quickly diverge in $p$-space. 
For the case of $x_0=0$, we have the general expression
\begin{equation}
A(t) = \frac{1}{\sqrt{\cosh(\tilde{\omega} t)}}
\exp\left[
\left(
\frac{p_0^2}{2m\tilde{\omega} h}
\right)
\left\{
\frac{\cosh(\tilde{\omega} t)-1 + i\sinh(\tilde{\omega}t)(2\cosh(\tilde{\omega} t) -1)}
{\cosh(\tilde{\omega} t)(\cosh(\tilde{\omega} t) - i\sinh(\tilde{\omega} t))}
\right\}
\right]
\end{equation}
In the limit when $t >> 1/\tilde{\omega}$, the hyperbolic functions
both approach $\exp(\tilde{\omega}t)/2$ and we have the limiting case
\begin{equation}
A(t>>1/\tilde{\omega}) 
\quad 
\longrightarrow
\quad
\frac{1}{\sqrt{\exp(\tilde{\omega}t)/2}}
\exp\left[- \frac{p_0^2}{2m\tilde{\omega} \hbar}(1-i)\right]
\end{equation}
The exponential (`dynamical') suppression once again is seen to 
`saturate', as in the free-particle case, and for the same reason, namely 
that both $x(t) - x_0$ and $\Delta x_t$ have the same large $t$ 
(here exponential) behavior. 
The resulting modulus is given by
\begin{equation}
|A(t)|^2
\longrightarrow
2e^{-\tilde{\omega} t}
\exp\left[
- \frac{p_0^2}{m\tilde{\omega} \hbar}
\right]
\end{equation}
which still becomes exponentially small, but now 
due to the (`dispersive') prefactor.
If one also has $x_0 \neq 0$, the expression above includes an
additional factor of $\exp(-x_0^2/\beta_0^2)$ (similar to that
in Eqn.~(\ref{sho_case}),  with no cross-term involving 
$x_0$ times  $p_0$.

\section{\label{sec_conclusions} Conclusions and discussion}

We have evaluated the autocorrelation function for time-dependent
Gaussian wave packet solutions for four quite different quantum 
mechanical systems, examining the behavior of $A(t)$ in terms of
both classical analogs and the quantum mechanical time-evolution of
the wave function magnitude and phases in both position- and momentum-space.
We have focused attention on the different contributions to the suppression
in $A(t)$ arising from `dispersive' (prefactor) and more `dynamical' 
(typically exponential) effects (which depend on the initial conditions) 
which may or may not saturate to a small but finite value, depending on 
the relationship between the classical dynamical behavior 
($x(t)$ or $\langle x \rangle _t$) and the quantum
mechanical spreading ($\Delta x_t$). We have extended earlier results
of Baltz \cite{baltz} in terms of a now standard analysis tool, namely the
autocorrelation function, directly comparing the behavior of $A(t)$ for 
several distinct classes of classical
behavior. Using the free-particle solution, we 
have also been able to exhibit a useful test case for quite general
theorems on the time-development of isolated quantum  systems, while
providing other non-trivial examples of closed-form results for $A(t)$
for Gaussian solutions. Additional examples which are simple extensions
of the results presented here are multi-dimensional free-particle,
uniformly accelerated particle, or harmonic oscillator solutions, 
where the autocorrelation function factorizes as $A(t) = A_{x}(t) 
\cdot A_{y}(t)$, or the related problem of a Gaussian
wave packet in a uniform magnetic field which admits Gaussian solutions
corresponding to classical circular  orbits.

\end{document}